# Realising high-dimensional quantum entanglement with orbital angular momentum

Melanie McLaren[1,2], Filippus S. Roux[2] and Andrew Forbes[1,2]

[1.] Laser Research Institute, University of Stellenbosch, Stellenbosch 7602, South Africa

[2.] CSIR National Laser Centre, P.O. Box 395, Pretoria 0001, South Africa

## Abstract

We report the first quantum entanglement experiment in South Africa. The spatial modes of the entangled photon pair are investigated with their potential for high-dimensional entanglement. The generation, measurement and characterisation of the entangled states are examined in detail, where we show high-dimensional entanglement in a Hilbert space of dimension 25. We highlight the experimental challenges contained within each step and provide practical techniques for future experiments in the quantum regime.

## Introduction

One of the most astonishing features of quantum mechanics is that of the entanglement of particles. First introduced as an objection to quantum mechanics by the famous Einstein, Podolsky and Rosen (EPR) thought experiment[1], entanglement represents the notion of non-local quantum correlations between two or more quantum-mechanical systems. That is, for an entangled pair of particles the measurement of an observable for one particle immediately determines the corresponding value for the other particle, regardless of the distance between the two particles.

This property of entangled systems led to a number of implications that disturbed many scientists and resulted in the emergence of hidden variable theories. Local hidden variable theory assumes that nature can be described by local processes, where information and correlations propagate at most at the speed of light and where the observables of a physical system are determined by some unknown (hidden) variables.

It was not until the 1960s, when Bell's inequality and its generalization, Clauser-Horne-Shimony-Holt (CHSH) Bell's inequality[2], demonstrated the possibility of practical experiments to test the validity of quantum theory with respect to local hidden variable theories. A slew of experiments followed to test Bell's inequality, each of which violated the inequality and in turn verified quantum mechanical predictions of entanglement[3,4,5,6]. These results encouraged the search for a method of producing maximally entangled states. Spontaneous parametric down-conversion (SPDC) has proved to be the most efficient technique in generating two-particle entanglement. Shih and Alley[7] were the first to demonstrate a violation of Bell's inequality using SPDC generated photon pairs. This was the start of various polarisation-entanglement experiments, however, in 2001 it was shown that orbital angular momentum (OAM) could also be used as a basis for entanglement and provided an advantage over polarisation with regard to the number of states available[8].

Quantum entanglement has sparked an interest in a number of scientific fields, such as quantum information processing[9], quantum cryptography[10] and quantum teleportation[11].

In this paper we report on the first entanglement experiment in South Africa. We create a bi-photon pair entangled in their spatial modes, allowing quantum states of high-dimension. In particular we illustrate how to generate, measure and quantify quantum entanglement of photons. Each process requires a number of sensitive steps, which we outline in detail. The paper should serve as a useful guide to encourage further quantum experiments in the region.

## Theory

### Entangled states

An entangled state can be simply viewed as two states which are inseparable. If we consider two subsystems A and B of a pure state $|\psi\rangle$, then that state can be written as

$$|\psi\rangle = |i\rangle_A |j\rangle_B, \qquad (1)$$

where $\{|i\rangle\}$ and $\{|j\rangle\}$ are orthonormal bases of A and B, respectively. This state is not entangled; it is merely a product of the two systems, allowing the properties of each system to be separated. A superposition state of the form of Eq. (1) is given as,

$$|\psi\rangle = \sum_{i,j} c_{ij} |i\rangle_A |j\rangle_B.  \qquad (2)$$

Here $c_{ij}$ represents the expansion coefficients. If the state cannot be separated into a product of the two systems A and B, it is entangled. For example, the following superposition state is entangled:

$$|\psi\rangle = \frac{1}{\sqrt{2}} \left( |i\rangle_A |j\rangle_B \pm |j\rangle_A |i\rangle_B \right). \qquad (3)$$

Both Eqs (2) and (3) represent a two-photon state, where each photon can carry bits of information, the number of which depends on the basis used. The degree of entanglement of any quantum system can be characterized by calculating the two-photon density matrix:

$$\rho = |\psi\rangle\langle\psi|. \qquad (4)$$

In order for this matrix to be physically allowed, it must have positive eigenvalues and a trace equal to unity. Eq. (4) is only true for pure quantum states. The density matrix has been determined for mixed states[12] and can be written as

$$\rho = \sum_n |\psi_n\rangle\langle\psi_n|. \qquad (5)$$

The density matrix allows various measurements to be calculated, which describe the quantum state, such as linear entropy, fidelity and concurrence. These will be defined in a subsequent section.

The density matrix is not limited to a two-dimensional qubit system, but can also be calculated for *d*-dimensional, qu*d*it, quantum systems. A qubit system can carry at most only two bits of information per photon. For example, polarization entangled photons can carry a horizontal or vertical state, represented by $|0\rangle$ and $|1\rangle$. A qu*d*it system can carry *d* bits of information per photon; an obvious benefit for quantum information processing applications[13].

## Angular momentum

In demonstrating quantum entanglement, a measurement must be made of one of the properties of single photons. It is possible to measure the position, momentum, energy and time of arrival of single photons[14-16]. However, the most well documented entanglement measurements have been demonstrated using angular momentum: both spin angular momentum (SAM) and orbital angular momentum (OAM).

**Spin angular momentum**

Angular momentum associated with circularly polarised light is known as spin angular momentum and is quantified by $\hbar$ per photon. The direction of the electric field oscillation of light as it propagates, specifies the type of polarisation. For linearly polarised light, the field oscillates in a single plane, whereas the field rotates about the propagation axis for circularly polarised light. The direction of the rotation specifies the handedness of the circular polarisation; clockwise specifies right-handed, anti-clockwise specifies left-handed.

Polarization offered an efficient way in which to demonstrate photon entanglement and much was learnt in terms of optimising the efficiency of generating and detecting entangled photons[17,18]. However, the limit on the amount of SAM carried per photon, prevented measurements of high-dimensional entanglement.

**Orbital angular momentum**

The OAM of light is associated with the spatial distribution of the light wave. In 1992 Allen et al.[19] demonstrated that laser beams with OAM have helical phase fronts and possess an azimuthal phase dependence of $\exp(i\ell\phi)$, where $\ell$ (azimuthal phase index of integer value) represents the number of azimuthal phase rotations in one full cycle from 0 to $2\pi$. A light beam propagating along z with an $\ell$-dependent azimuthal phase has a field amplitude described by,

$$\psi(r,\phi,z) = \psi_0(r,z)\exp(i\ell\phi), \qquad (6)$$

where $\psi_0$ is an amplitude distribution, $\ell$ is an integer and $\phi$ is the azimuthal angle. A common example of such beams is the Laguerre-Gaussian (LG) modes, a complete basis set of orthonormal modes, shown in Eq. (7).

$$LG_{p\ell} = \sqrt{\frac{2p!}{\pi(p+|\ell|)!}} \frac{1}{w(z)} \left[\frac{r\sqrt{2}}{w(z)}\right]^{|\ell|} \exp\left[\frac{-r^2}{w^2(z)}\right] L_p^{|\ell|}\left(\frac{2r^2}{w^2(z)}\right) \exp[i\ell\phi] \times$$
$$\exp\left[\frac{ik_0 r^2 z}{2(z^2 + z_R^2)}\right] \exp\left[-i(2p+|\ell|+1)\tan^{-1}\left(\frac{z}{z_R}\right)\right] \qquad (7)$$

Here $\ell$ and $p$ are the azimuthal and radial mode index, respectively. The 1/e radius of the Gaussian term is given by $w(z) = w(0)\left[(z^2 + z_R^2)/z_R^2\right]^{1/2}$ with the beam waist $w(0)$, the Rayleigh range by $z_R$, the Guoy phase by $\tan^{-1}(z/z_R)$ and $L_p^{|\ell|}(x)$ represents a Laguerre polynomial. The intensity distribution of an LG mode with $\ell > 0$, consists of a zero on-axis intensity surrounded by $p+1$ concentric rings. The radius of maximum intensity for such modes is proportional to $\sqrt{\ell}$ for $p = 0$.

This non-zero OAM results from the helical phase front of LG beams. This is not a unique property to LG beams and is found in both higher-order Bessel beams[20] and Ince-Gaussian beams[21]. It is therefore possible to study OAM entanglement using modes other than those in the LG basis.

An analogy between polarisation and OAM can be made using the Poincare sphere and its equivalent, the Bloch sphere[22]. Any polarisation state can be represented in a convenient graphical manner using the Poincaré sphere [Fig. (1)].

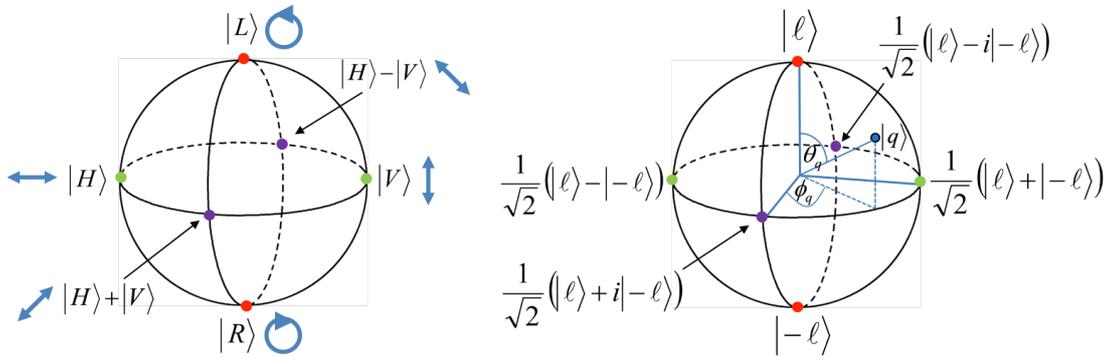

Figure 1: Analogy between polarization and OAM can be seen using (a) the Poincare sphere and (b) the Bloch sphere.

According to this representation, the two circularly polarised states lie at the north (right-handed) and south (left-handed) poles. Linearly polarised states lie on the equator, while the states which lie between the poles and equator represent elliptical polarisation. Similarly, any OAM state can be represented on the surface of an equivalent sphere, known as the Bloch sphere [Fig. 1(b)]. The north and south poles of the sphere represent the states $|\ell\rangle$ and $|-\ell\rangle$, while the points around the equator, $\theta = \pi/2$, represent the superposition states. Any vector $|q\rangle$ on the sphere can be described using the spherical co-ordinates $\theta$ and $\phi$ by,

$$|q\rangle = \cos\left(\frac{\theta}{2}\right)|\ell\rangle + \exp(i\phi)\sin\left(\frac{\theta}{2}\right)|-\ell\rangle. \qquad (8)$$

Figure (2) shows superpositions of OAM modes in two different bases: the LG basis and the Bessel-Gauss (BG) basis.

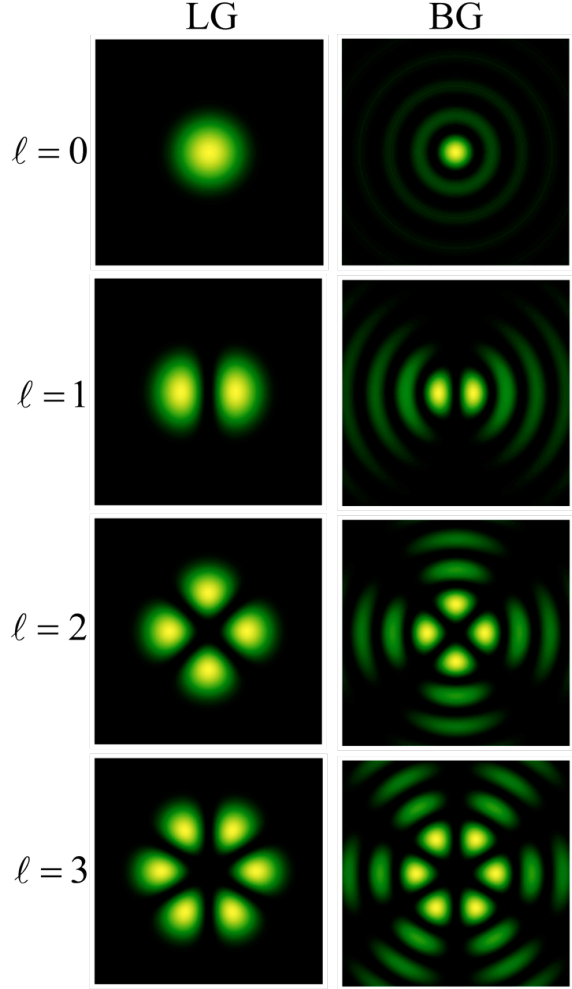

Figure 2: Intensity distributions for superposition modes of OAM for different azimuthal indices. For each image, the superposition is between $\ell$ and $-\ell$. Both bases, LG and BG, can be used to measure OAM. The higher order Bessel modes vary with the azimuthal phase index[23].

**Entangled states in OAM**

The seminal paper by Mair et al.[8] was the first experiment to demonstrate OAM as a property of single photons produced by SPDC. They showed that OAM is conserved in the SPDC process and consequently showed entanglement involving these modes[24]. That is, the OAM of the entangled photon pair must sum to the OAM of the pump. The two-photon state for OAM can be written as

$$|\Psi\rangle = \sum_\ell a_{\ell,-\ell} |\ell\rangle |-\ell\rangle, \qquad (9)$$

where $|a_\ell|^2$ is the probability of finding one photon in state $|\ell\rangle$ and the other in state $|-\ell\rangle$, where the pump beam has zero OAM. As $\ell$ can assume any integer value, Eq.

(9) is true for *d*–dimensional two-photon states, where $\ell$ ranges over *d* different values.

The conjugate variable of OAM is angular position, which can be described by an aperture with an angular width $\phi$. Similar to Heisenberg's uncertainty relationship between momentum and position, the relationship between OAM and angular position is described by[25],

$$[\Delta(\ell\hbar)][\Delta(\phi)] \geq \frac{\hbar^2}{4}. \qquad (10)$$

A violation of the inequality in Eq. (10) satisfies the EPR-Reid criterion[26], which is analogous to the original EPR paradox. These correlations demonstrate entanglement not only for discrete variables such as OAM but for continuous variables like angular position as well.

## Generation of entangled photons

The most commonly used and most efficient method of producing entangled photon pairs is that of spontaneous parametric down conversion (SPDC)[27]. This non-linear optical process decays a pump photon into two photons (signal and idler) in a crystal of optical non-linearity, $\chi^2$. Both energy and momentum are conserved in this decay process, also known as the phase matching conditions:

$$\omega_p = \omega_s + \omega_i \qquad (11)$$

$$\vec{k}_p = \vec{k}_s + \vec{k}_i \qquad (12)$$

Here, $\omega_p, \omega_s, \omega_i$ are the frequencies and $\vec{k}_p, \vec{k}_s, \vec{k}_i$ the wave vectors of the pump, signal and idler photon, respectively.

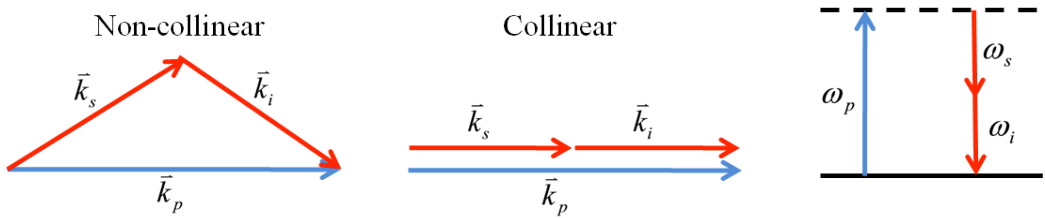

Figure 3: Phase matching conditions for non-collinear and collinear SPDC.

Due to these conditions, the measurement of one photon in a particular direction and energy, forces the existence of the other correlated photon pair of definite energy and direction. There are two types of SPDC, namely type I and type II. In type I, the down-converted photons are produced with the same polarisation, orthogonal to that of the pump. Photons of the same wavelength are emitted on concentric cones centred around the pump axis of propagation. The diameter of the cone depends on the angle between the pump beam and the optical axis of the crystal. Type II SPDC emits one photon with the same polarisation as the pump and the other with orthogonal polarisation. In both cases, the process is said to be degenerate if the down-converted photon pair have the same wavelength (i.e. $\lambda_i = \lambda_s = 2\lambda_p$) and non-degenerate otherwise. SPDC can be further characterised into collinear and non-collinear processes, depending on the position of the optic axis relative to the pump. By simply changing the tilt of the crystal, we are able to change the opening angle of SPDC, moving from non-collinear to collinear down-conversion [see Fig. (4)], allowing an optimal position to be found depending on the application.

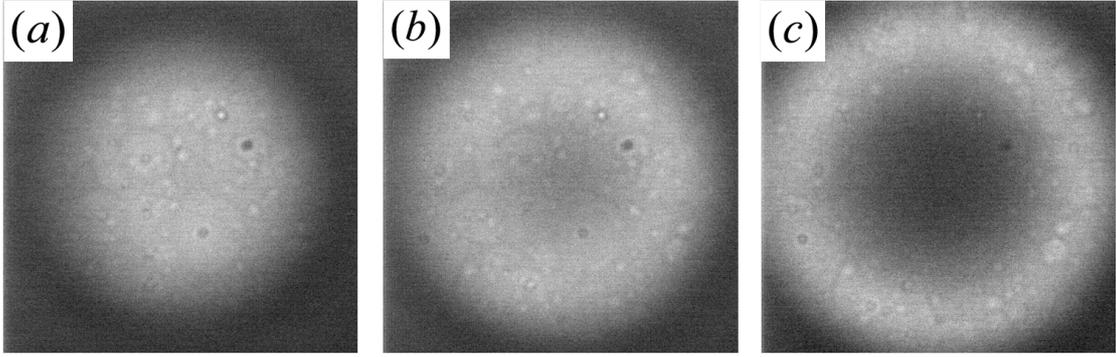

Figure 4: SPDC is the most efficient method for producing entangled photons, however, the probability of a spontaneous decay into a pair of entangled photons is very low, approximately 1 in every $10^{12}$ photons are down-converted. Therefore a very sensitive electron multiplier CCD camera is needed to image the ring of photons. (a) Far-field image of the collinear down-converted light from the BBO crystal. (b) Far-field image of the near-collinear down-converted light from the BBO crystal. (c) Far-field image of the non-collinear down-converted light from the BBO crystal. The change from non-collinear to collinear requires a very small change in tilt.

It has been found that the intensity profile of the far-field down-converted light follows the function[28]

$$I(r) = \text{sinc}^2\left(\frac{ar^2}{f^2} + \alpha\right), \qquad (13)$$

where $f$ is the focal length of the Fourier lens with radial coordinate $r$. The phase-matching parameter establishes the opening angle of SPDC and is given by $\alpha = (|\vec{k}_p| - |\vec{k}_s| - |\vec{k}_i|)L/2$, and $a = (|\vec{k}_s| + |\vec{k}_i|)L/4n^2$, for refractive index $n$, crystal length $L$. The propagation of light through an optical system can be described by étendue, which when normalised to the wavelength $\lambda$, gives an estimate of the number of detectable transverse modes, written as[29]

$$N = \frac{A\Omega}{\lambda^2}. \qquad (14)$$

Here, $A$ is the area of the near-field beam and $\Omega$ is the far-field opening angle. We have chosen to measure OAM entanglement in the near-field, effectively setting $A$ to be constant. Thus, as we change $\alpha$ from collinear to near-collinear, the opening angle $\Omega$ increases, yielding a larger OAM spectrum.

## Spiral bandwidth

We know that SPDC produces pairs of photons that are entangled in OAM[8]. The number of OAM modes generated in this process is known as the generation spiral bandwidth and is dependent on two main factors: the size of the pump beam and the length of the crystal. As the radial intensity distribution increases with $\sqrt{\ell}$, the size of the pump waist determines the highest OAM mode achievable. Thus a larger pump waist is favourable over a tightly focused spot. The phase matching conditions become more restrictive for thicker crystals, which consequently diminishes the probability of generating higher OAM modes. A thinner crystal (e.g. 0.5 mm thick) will produce a wider spectrum of OAM modes, but at the cost of efficiency. A compromise between the spiral bandwidth and efficiency will be determined by the aim of the experiment.

The bandwidth generated at the crystal is not necessarily the bandwidth detected. This is dependent on the design of the experiment, including the single-photon source, the measurement methods and the detection components. The following section describes each aspect of the experimental setup used to demonstrate quantum entanglement.

## Experimental design

### Components

Figure 5 shows a general idea of the design of the experiment. A source to generate single photon pairs is an obvious requirement. As previously mentioned, SPDC is a widely used method, which we too use for this paper. A particular property of a photon needs to be chosen so as to determine the measurement scheme. Such properties include the measurement of the conjugate variables energy and time, or momentum and position. We chose to measure the OAM of photons mainly owing to its potential for high dimensional entanglement and the ease at which we can measure OAM using SLMs. Finally, we also need a method to detect not only single photons but also the coincidence counts from entangled photon pairs.

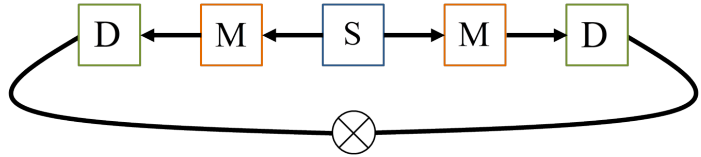

Figure 5: Schematic of the experimental setup used to generate, measure and detect entangled photons. The source, S, generates the photon pairs, which are then measured in a particular basis at M and then finally detected at D.

We examine these experimental segments to highlight the challenges involved in setting up each.

**Source**

The SPDC process requires a balance between generating sufficient photon pairs for reasonable counting times, and errors introduced when many photon pairs are produced within the same time window. The balance between these parameters is largely determined by the energy and the repetition rate of the pump source. In our experiment a mode-locked ultraviolet pump source with a wavelength of 355 nm and average power of 350 mW was used. The laser produces pulses at 80 MHz, each pulse made up of $\sim 10^9$ photons. The SPDC process produces on average 1 photon pair in every $\sim 10^5$ pulses, or 800 per second. This relates to an efficiency of $\sim 10^{-12}$. The time between photon pairs in turn dictates the time window of the detection system (gating time). Since the pulses arrive at intervals of 12.5 nanoseconds, we select a gating time of 12.5 nanoseconds to minimise the error of multiple photon pairs arriving at the same time. The high repetition rate of the laser is only possible with mode-locked systems, and is used here only as a means to increase the sample rate. Increasing the energy of the pulse would also increase the pair production rate through an improved efficiency of the SPDC process. Unfortunately this also increases the error in the system by way of multiple pair generation. The length of the non-linear crystal also has an effect on the SPDC efficiency, that is, a thicker crystal will produce more photon pairs per pulse. We observed an increase in coincidence counts when changing from a 0.5-mm-thick crystal to a 3-mm-think crystal of approximately three times. However, this must be weighed up against the generated spiral bandwidth (the width of the distribution of OAM modes), which is also dependent upon crystal length.

**Measurement**

A spatial light modulator enables the phase of an incoming beam to be shaped according to the encoded hologram. That is, a Gaussian beam illuminating a phase-only forked hologram of particular azimuthal index, $\ell$, produces a helically-phased beam in the first diffraction order. This process also operates in reverse, in that a beam with OAM $\ell$ incident on a forked hologram with an azimuthal index $-\ell$, will produce a Gaussian beam.

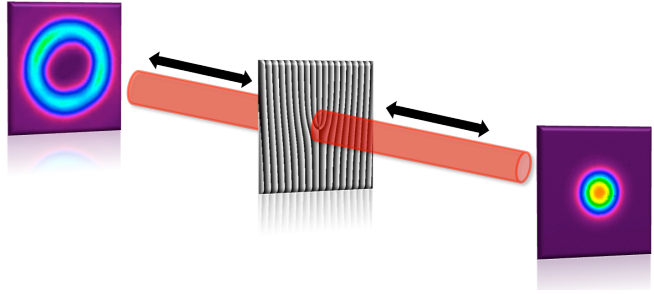

Figure 6: A spatial light modulator encoded with an azimuthal phase dependence, $\exp(i\ell\phi)$, shapes a Gaussian beam into a helically-phased beam in the first diffraction order. This process also works in reverse such that an LG beam can be converted into a Gaussian beam.

Only the fundamental mode (a Gaussian beam) can propagate through single-mode fibres (SMF). The hologram on the SLM together with the SMF, act as a "match-filter" such that an incoming beam with OAM $\ell$ will only couple into the SMF if the hologram is encoded with the same azimuthal phase index, $\ell$.

The angular position states are measured by defining a narrow "slice" of the hologram[30], representing an aperture [see Fig. (7)]. The orientation and width of the aperture can be varied; however, a very narrow slice results in a significant reduction in the number of coincidence counts detected. Thus, there is a minimum limit to which the width can be defined.

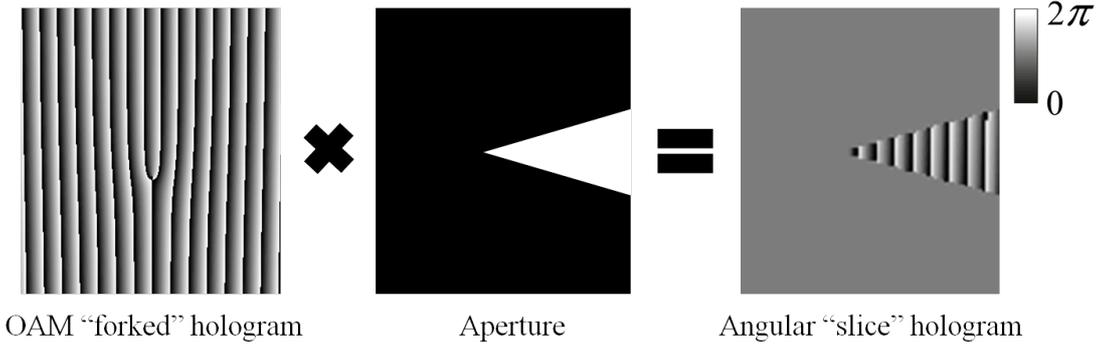

Figure 7: Hologram used to measure angular position correlations. The conjugate variable of OAM is angular position. By setting a particular aperture, we can measure the angular position correlations by varying the aperture width and its orientation.

**Detection**

The detection of photon pairs requires particular devices, which are both electronically fast and efficient. The single photons are typically measured using avalanche photo-diodes (APDs) with a quantum efficiency of approximately 60%. Evidently, the laboratory must be as dark as possible when using single photon detectors. Nonetheless, the measured dark count rate, the single photon count rate of the dark laboratory without lasing source, is approximately $200\,\text{s}^{-1}$ as the laboratory does allow in some external light.

The coincidence count rate is recorded using a coincidence counter connected to two single photon detectors. When a photon is detected by one detector it starts the trigger in the second detector and when a photon arrives in the second, a coincidence is recorded. Of course, there must be a limit placed on the interval in which the second photon is detected, known as the gating time. There are a number of counting instruments available, which vary in price and gating time. The HydraHarp 400 was used with a gating time set at 12.5 ns. Even with a very narrow time interval, uncorrelated photon coincidence counts may be recorded, know as accidental counts and can be estimated as $C_{acc} = S_1 S_2 \Delta t$, where $S_1$ and $S_2$ are the single count rates from each detector and $\Delta t$ is the gating time.

# Experiment

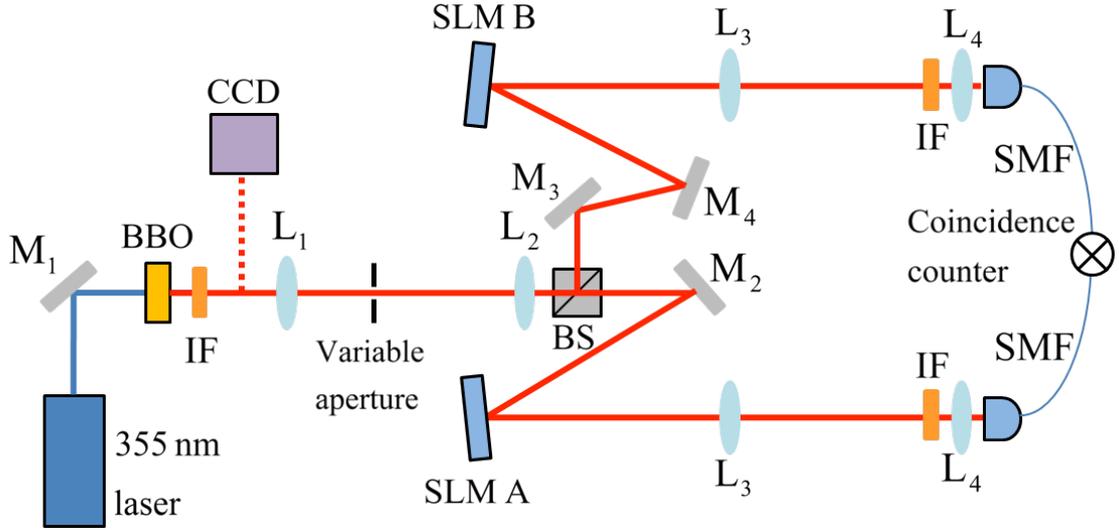

Figure 8: Experimental setup used to detect the OAM eigenstate after SPDC. The plane of the crystal was relay imaged onto two separate SLMs using lenses, $L_1$ and $L_2$ ($f_1$ = 200 mm and $f_2$ = 400 mm), where the LG modes were selected. Lenses $L_3$ and $L_4$ ($f_3$ = 500 mm and $f_4$ = 2 mm) were used to relay image the SLM planes through 10 nm band-width interference filters (IF) to the inputs of the single-mode fibres (SMF).

Figure 8 shows our experimental setup. A mode-locked ultraviolet pump source with a wavelength of 355 nm and average power of 350 mW was used to pump a 3-mm-thick type I barium borate (BBO) crystal to produce collinear, degenerate entangled photon pairs via SPDC. An interference filter was placed after the crystal to reflect the pump beam and transmit the 710 nm down-converted light. The front plane of the crystal was then imaged ($f_1$ = 200 mm, $f_2$ = 400 mm) onto two separate phase-only spatial light modulators (SLMs). Just as polarisers were used to "select" a particular polarization state, the SLMs allow a specific state to be chosen into which the photon will be projected. Initially, the Laguerre-Gaussian basis set was chosen to measure the OAM states, however any orthogonal basis set can be chosen such as the Bessel-Gaussian basis[23]. The SLM planes were then reimaged ($f_3$ = 500 mm, $f_4$ = 2 mm) and coupled into single-mode fibres (mode-field diameter = 4.6 $\mu$m) so as to extract only the Gaussian modal components. Interference filters centred at 710 nm were placed in front of each fibre coupler to prevent any scattered pump light to enter the fibres. The fibres connected to single-photon avalanche detectors allowed the arrival of a photon pair to be registered using a HydraHarp 400 coincidence counter.

Table 1: Specifications of the experimental equipment used to perform quantum entanglement.

| Equipment | Manufacturer | Specifications |
|---|---|---|
| **Laser source** | Newport | 355 nm, 350 mW |
| **Non-linear crystal** | Castech | BBO, type I, degenerate |
| **Spatial light modulator** | HoloEye | Phase-only, NIR, 1920x1080 pixels |
| **Interference filters** | Thorlabs | Central wavelength: 710 +/- 10 nm |
| **Single-mode fibre** | Thorlabs | 630-680 nm |
| **Avalanche photodiodes** | Perkin Elmer | Dark count ~ 200 counts |
| **Coincidence counter** | PicoQuant: HydraHarp400 | 8 channels |

An object can be imaged in two ways, either using a single lens (2f-imaging system) between the object and image plane or using two lenses (4f-imanging system). While both methods image the intensity of the object, only the latter method images both the intensity and phase of the object. This is important when considering the properties of two-photon correlations[31], as the wave fronts of the signal and idler beams must sum to that of the pump beam in order to detect a coincidence signal. An additional lens function can be added to the SLM to compensate for a mismatch in wave fronts. However this method requires additional optimisation for each SLM. A simpler and more effective method, which we use in our experimental setup, makes use of a 4f-imaging system. Assuming the pump beam has a planar wave front at the crystal, a 4f-imaging scheme will image modes with planar wave fronts onto both SLMs.

As the down-conversion process is not very efficient, alignment of the system becomes rather difficult. We used a method of "back projection" or "retrodiction", first proposed by Klyshko[32], to align the system, by passing a 710 nm diode laser beam through each of the single mode fibres so as to pass through the system in reverse. There are a number of benefits in using this method.

Firstly, the pump beam and both back-projected beams must overlap at the plane of the crystal. Initially the SLMs were switched off and therefore only acted as mirrors, allowing a single beam to be aligned with the pump beam. Thereafter, the SLMs were switched on with a grating hologram to separate the diffraction orders. The tilt of the SLMs was changed such that the first diffraction order from each SLM was overlapped with the pump beam. The measured coincidences can also be numerically determined by calculating the overlap integral of the back-projected signal, idler beams with the pump field at the crystal plane. The coincidence rate is proportional to the overlap intergal[33], that is,

$$C \propto \frac{\left|\int \Psi_s^* \Psi_i^* \Psi_p dA\right|^2}{\sqrt{\int |\Psi_s^* \Psi_p|^2 dA \int |\Psi_i^* \Psi_p|^2 dA}}, \quad (15)$$

where $\Psi_s$, $\Psi_i$ and $\Psi_p$ are the modes of the signal, idler and pump beams, respectively. If a back-projected beam of $\ell = 0$ is of equal size with the pump, then the overlap for $\ell = 0$ will be very strong, while higher OAM modes, whose intensity distribution increases with $\sqrt{\ell}$, will have very weak overlaps. Thus, if the back-projected beam of $\ell = 0$ is smaller than the pump, the overlap will be less strong for the $\ell = 0$ mode but also more significant in higher modes. The ratio of the pump waist $w_p$ with the signal (idler) waist $w_{s,i}$ is given as[34]:

$$\gamma_{s,i} = \frac{w_p}{w_{s,i}}. \quad (16)$$

In terms of the imaging systems shown in Fig. (8), we calculated our ratio to be $\gamma = 2$. This appears to be a fair ratio when finding a balance between efficiency and spiral bandwidth. Miatto et al.[34] investigate three different ratios; 0.5, 2 and 4 with regard to LG modes with $p = 0$. They show that although the spiral bandwidth does increase with $\gamma$, the count rate for $\gamma = 4$ drops almost to the accidental count rate.

Back-projection also allows us to view the modes that we generate on the SLMs, which guarantees that we are indeed measuring in the correct basis. We placed a flip-up mirror between the BBO crystal and lens $L_1$ so as to reflect the back-projected light onto a CCD camera (placed 200 mm from $L_1$). The alignment of the beam with the centre of the hologram is very important, as seen in Fig. (9), where a misaligned beam results in an uneven intensity distribution.

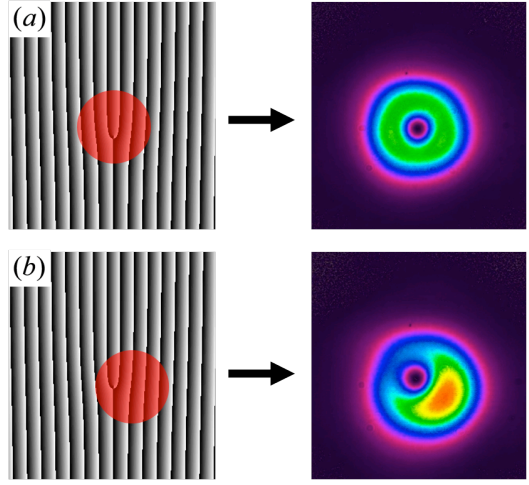

Figure 9: The alignment of the beam with the centre of the SLM has a significant effect on the coincidence measurements, such that measurements with a theoretically value of zero become non-zero with misalignment.

On the single photon level, this uneven distribution corresponds to measuring non-zero coincidence counts where there should be none. Back-projection provides visual confirmation that the beam is centred on the hologram. However, we performed an additional optimisation on the spatial position of the beam on the SLM using the knowledge that OAM is conserved in down-conversion. That is, for a pump beam with OAM of $\ell = 0$, we would only expect to see a coincidence count when $\ell_s + \ell_i = 0$. Thus, if a hologram of $\ell = 0$ is encoded onto SLM A and $\ell = 1$ on SLM B, we do not expect to observe any coincidence counts. If we do see counts, the $\ell = 1$ hologram is repositioned until a minimum count rate is obtained and then repeated for $\ell = 1$ on SLM A.

The measurements of OAM are sensitive to lateral position alignment, while the angular position measurements are sensitive to the axial positions of the image planes. Observing the back-projected beam on the CCD allowed us to ensure both SLMs were placed in the correct image plane. Both lenses $L_1$ and $L_2$ were placed upon micrometer translation stages to allow for small adjustments along the axis of propagation to find the correct image plane. The size of variable aperture between lenses $L_1$ and $L_2$ was decreased until a clear image of the "slice" was seen.

Once each step was carefully executed, both single-mode fibres were connected to their respective detectors and the flip-up mirror was moved out of the beam path. With the UV laser pumping the crystal, the coincidence count rate was recorded. A few minor adjustments were made to the angular position of both SLMs so as to maximise the coincidence count rate.

## Quantum measurements

After following a rigorous alignment procedure, we were able to perform a number of fundamental tests. The first of which was to demonstrate the conservation of OAM. Each SLM was encoded with holograms ranging $\ell$ from -20 to 20, one after another. Figure 10 shows the measured coincidence counts known commonly as the spiral bandwidth.

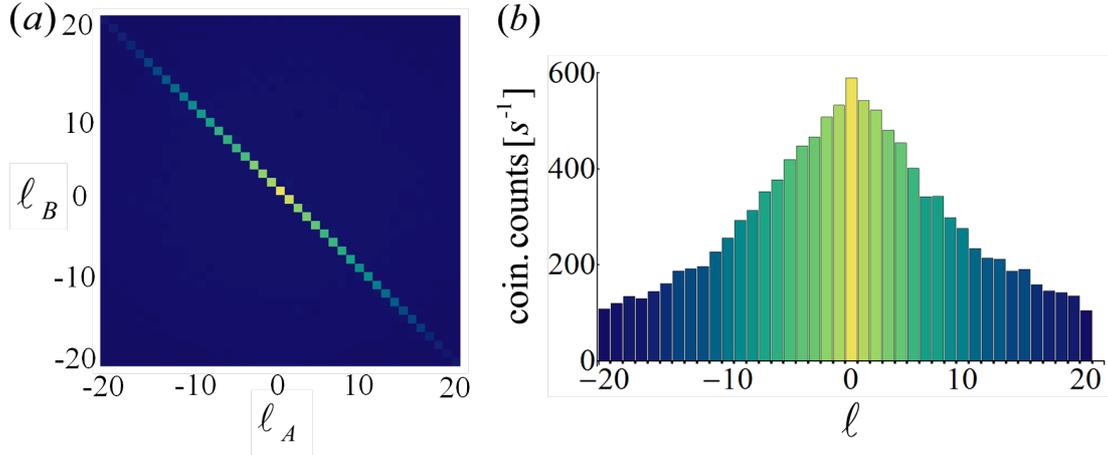

Figure 10: Experimental results showing (a) a density plot of the coincidence counts per second and (b) the non-zero diagonal elements representing a spiral bandwidth plot. The FWHM of the spiral bandwidth is approximately 15.

The anti-correlated diagonal is consistent with OAM conservation, that is $\ell_p = \ell_s + \ell_i$. While the coefficients in the OAM spectrum demonstrate a decreasing trend from $\ell = 0$, the size of the mode is another contributing factor as the mode increases with the azimuthal index, which results in a loss of efficiency, resulting in a decreasing trend from $\ell = 0$. Another important feature obtained from Fig. (10a) is the values of the off diagonal elements. Theoretically these should be zero, but experimentally this is often impossible to achieve as the spiral bandwidth is highly sensitive to misalignment. We measured the off-diagonal elements to be less than 5% of their corresponding diagonal element. From Fig. (10b) we measured the full-width-half-maximum (FWHM) value to be 15.

The spiral bandwidth experiment offers an effective method to test whether the setup is correctly aligned. Similarly, the angular position can be used to ensure the optics are placed in the correct image planes. An angular sector hologram was encoded onto each SLM; one hologram fixed at a particular orientation while the other hologram rotated in small increments through $2\pi$. In a ghost imaging experiment, an aperture is placed in one arm and the detector in the other arm is scanned through its transverse position, resulting in the reconstruction of the aperture shape. Similarly, the width of the angular hologram was determined from the measured coincidence counts. A sharp coincidence peak was recorded when the holograms were both orientated at the same angle, see Fig. (11), where the width of the peak gives the width of the angular "slice".

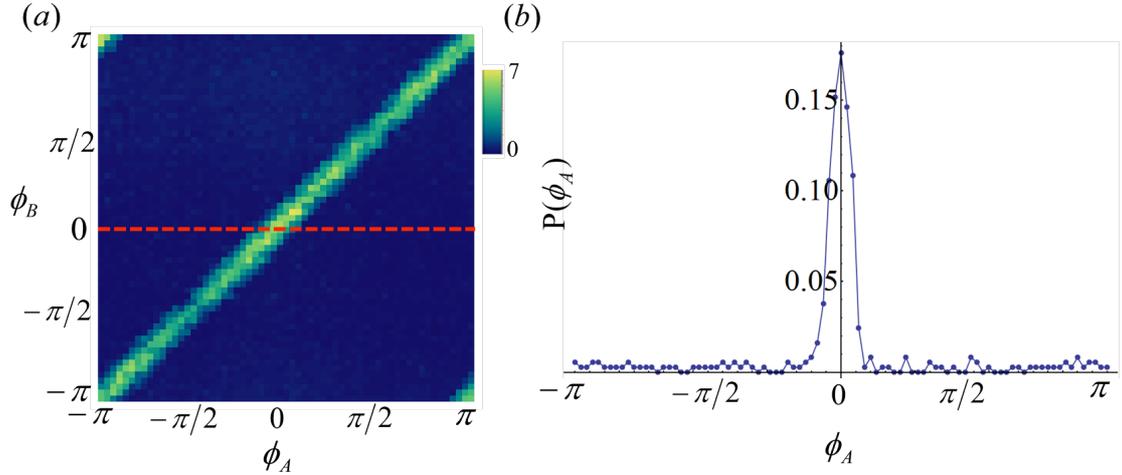

Figure 11: Experimental measurements showing (a) a density plot of the coincidence counts per second using the angular holograms. The holograms on each SLM were rotated through $2\pi$. (b) Probability distribution of the angular position $\phi_A$ for $\phi_B = 0$, taken along the red dotted line in (a).

From the data recorded for both the spiral and angular bandwidths, we calculated the uncertainty relationship between the two. A profile from the centre of each spectrum was plotted and fitted with a Gaussian distribution (Fig. (12)), which gave the following widths $[\Delta\ell]^2 = 0.128 \pm 0.023$ and $[\Delta\phi]^2 = 0.056 \pm 0.006$. Therefore, by taking the product of the two $[\Delta\ell]^2[\Delta\phi]^2 = 0.007 \pm 0.001$, the EPR-Reid criterion is violated as the product is clearly smaller than the uncertainty relation of 0.25 in Eq. (10).

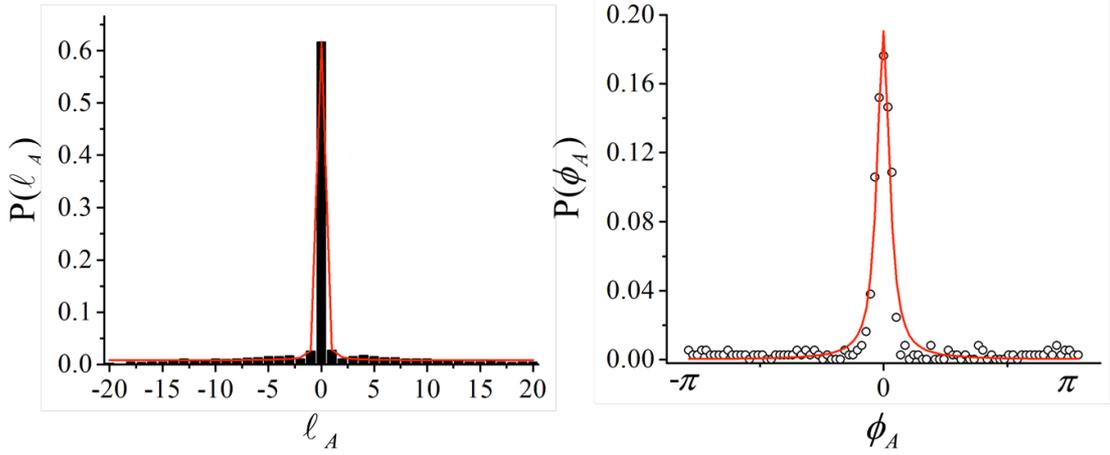

Figure 12: Probability distributions for (a) the OAM $\ell_A$ for $\ell_B = 0$ and (b) the angular position $\phi_A$ for $\phi_B = 0$. A Gaussian distribution has been fitted to both to determine the widths of each plot, which were used to demonstrate the EPR-Reid criterion.

**Bell inequalities**

The EPR paradox does not eliminate the possibility of hidden variables. To do this, a violation of Bell's inequality must be shown. In order to demonstrate this, the correlations between the two entangled photons must also be observed for superposition states[35], described by

$$|\Psi\rangle = \frac{1}{\sqrt{2}}\left(|\ell\rangle + \exp(i\ell\theta)|-\ell\rangle\right). \quad (17)$$

Here $\theta$ denotes the degree of rotation. By choosing a particular value for $\ell$, we generated superposition holograms for a range of angles $\theta$. The holograms were varied on both SLMs, by fixing one at orientation $\theta_A$ and rotating the other $\theta_B$, and the coincidence count rates were measured. Bell showed that the sinusoidal behaviour seen in figure B cannot be simulated by classical correlations and the deviation from classical theory can be calculated using Bell's inequality or a variation thereof derived by Clauser et al[2].

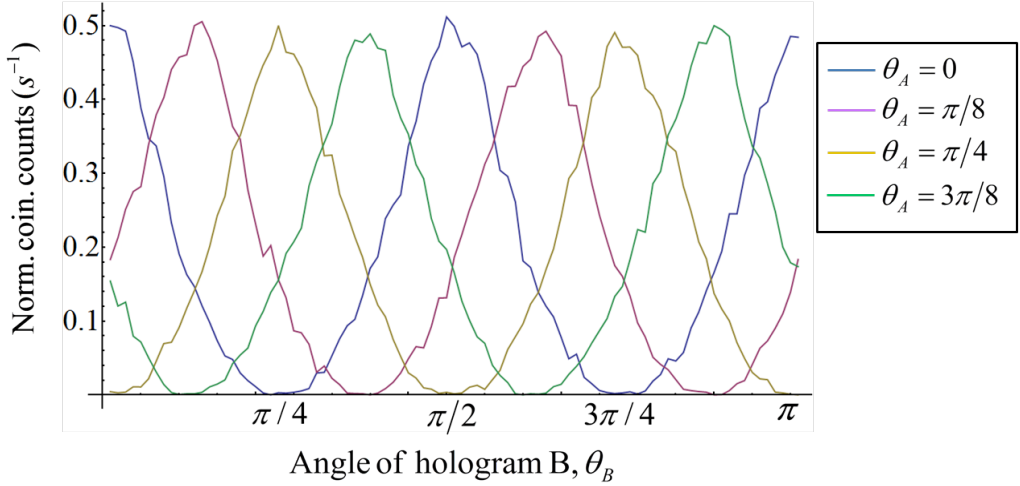

Figure 13: The normalised coincidence counts as a function of the orientation of the holograms on each SLM. The orientation of the hologram on SLM A was fixed while those on SLM B were rotated from 0 to $\pi$.

The Bell parameter S can be defined as[35]:

$$S = E(\theta_A, \theta_B) - E(\theta_A, \theta_B') + E(\theta_A', \theta_B) + E(\theta_A', \theta_B'). \qquad (18)$$

Where $\theta'$ is a different orientation from $\theta$. $E(\theta_A, \theta_B)$ is calculated directly from the measured coincidence counts $C(\theta_A, \theta_B)$ at particular orientations,

$$E(\theta_A, \theta_B) = \frac{C(\theta_A, \theta_B) + C\left(\theta_A + \frac{\pi}{2\ell}, \theta_B + \frac{\pi}{2\ell}\right) - C\left(\theta_A + \frac{\pi}{2\ell}, \theta_B\right) - C\left(\theta_A, \theta_B + \frac{\pi}{2\ell}\right)}{C(\theta_A, \theta_B) + C\left(\theta_A + \frac{\pi}{2\ell}, \theta_B + \frac{\pi}{2\ell}\right) + C\left(\theta_A + \frac{\pi}{2\ell}, \theta_B\right) + C\left(\theta_A, \theta_B + \frac{\pi}{2\ell}\right)}. \qquad (19)$$

The inequality is violated when $|S| > 2$. For the CHSH inequality, the upper limit for an entangled system is $|S| \leq 2\sqrt{2}$. For $|\ell| = 2$, we observed a violation of the inequality by 26 standard deviations, $S = 2.78 \pm 0.03$, indicating an entangled system.

## State tomography

Lastly, it is important to form a characterisation of the entangled states by measuring the degree of entanglement. This is attained by performing a state tomography on the system. We previously introduced the concept of a density matrix used to describe the statistical state of a quantum system. Using particular coincidence measurements, we are able to reconstruct[36] the two-photon density matrix:

$$\rho = \frac{1}{4}\sigma_0 \otimes \sigma_0 + \sum_{mn} \rho_{mn} \sigma_m \otimes \sigma_n \qquad (20)$$

where $\sigma_x$ represents the Pauli matrices for $x = 1, 2, 3$. From the Bloch sphere[22], we focused on the equator states $\theta = 0, \pi/4, \pi/2, 3\pi/4$ in Eq. (16) together with the pure states $|\ell\rangle$ and $|-\ell\rangle$, shown in Fig. (14).

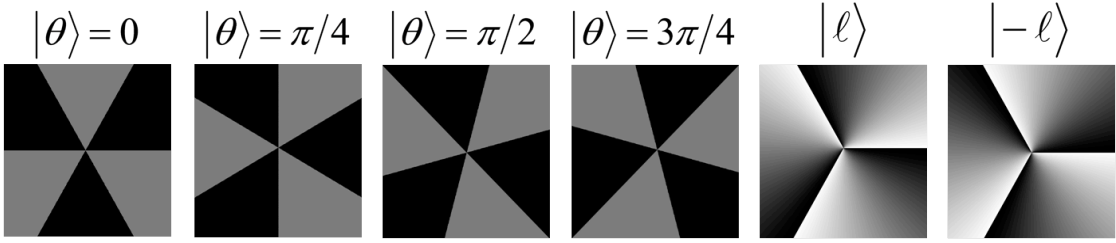

Figure 14: Holograms used to perform a state tomography on the system with the aim of calculating the density matrix. These include two pure states and the corresponding four superposition states. All six holograms were cycled through on each SLM, resulting in 36 coincidence measurements.

In matrix form, the qubit density matrix is written as:

$$\rho = \begin{pmatrix} a_{11} & a_{12}e^{i\phi_{12}} & a_{13}e^{i\phi_{13}} & a_{14}e^{i\phi_{14}} \\ a_{12}e^{i\phi_{12}} & a_{22} & a_{23}e^{i\phi_{23}} & a_{24}e^{i\phi_{24}} \\ a_{13}e^{i\phi_{13}} & a_{23}e^{i\phi_{23}} & a_{33} & a_{34}e^{i\phi_{34}} \\ a_{14}e^{i\phi_{14}} & a_{24}e^{i\phi_{24}} & a_{34}e^{i\phi_{34}} & a_{44} \end{pmatrix} \qquad (21)$$

Here $a_{ij}$ and $\phi_{ij}$ are the amplitudes and phases of the density matrix elements. The probability to detect one photon in state $|\ell\rangle$ and the other in state $|-\ell\rangle$ is expressed by the diagonal terms. The off-diagonal terms are determined from measurements in the superposition states.

We generated six different states on each SLM, resulting in a total of 36 coincidence measurements. Therefore for a two-dimensional state, we use an over-complete set of measurements (36) to determine the 16 density matrix elements. We used a least squares fitting program to calculate the best density matrix according to our measurements. We followed the procedure in Jack et al.[36] where the 10 amplitudes and 6 phases from the density matrix in Eq. (21) are chosen such that the following equation is minimised:

$$\chi^2 = \sum_{i=1}^{36} \left( \frac{C_i^M - C_i^P}{\sqrt{C_i^M + 1}} \right) \qquad (22)$$

The experimentally measured coincidence counts are represented by $C_i^M$ and the counts predicted from the density matrix are given by $C_i^P$. In matrix form, the two-dimensional reconstructed density matrix was calculated to be

$$\rho = \begin{pmatrix} 0.011 & -0.001 & 0 & -0.002 \\ -0.001 & 0.48 & 0.48 & 0.036 \\ 0 & 0.48 & 0.49 & 0.036 \\ -0.002 & 0.036 & 0.036 & 0.012 \end{pmatrix} + \begin{pmatrix} 0 & 0.043 & 0.048 & 0.003 \\ -0.043 & 0 & -0.039 & 0.042 \\ -0.048 & 0.039 & 0 & 0.048 \\ -0.003 & -0.042 & -0.048 & 0 \end{pmatrix}. \quad (23)$$

Eq. (23) shows the real and imaginary parts of the density matrix, shown graphically in Fig. (15).

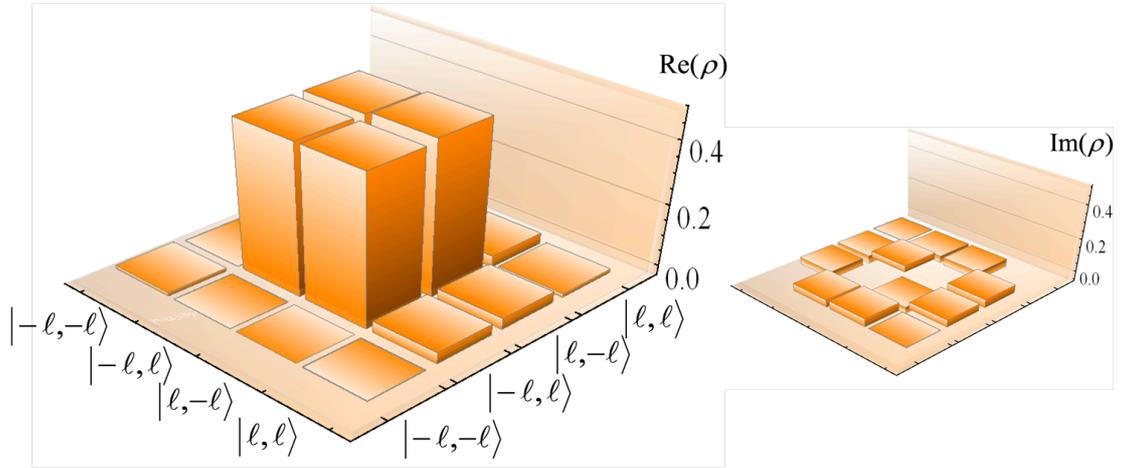

Figure 15: Graphically representation of the (a) real and (b) imaginary parts of the two-dimensional density matrix.

As previously mentioned, the density matrix can be used to calculate various quantifying parameters. Firstly, the linear entropy[37] $S_L = 4/3 \left[ 1 - \text{Tr}(\rho)^2 \right]$ defines the purity of the system, where the linear entropy of a pure entangled state is zero. We calculate a linear entropy of 0.02 for the above density matrix. The fidelity is a measure of how close our reconstructed state is to the target state, which is the (pure) maximally entangled state in this case. The fidelity is given by

$$F = \left[ \text{Tr} \left\{ \left( \sqrt{\rho_T} \rho_d \sqrt{\rho_T} \right)^{1/2} \right\}^2 \right]. \quad (24)$$

Here, $\rho_T = |\psi_T\rangle\langle\psi_T|$ is the target state with $|\psi_T\rangle = 2^{-1/2} \left( |\ell\rangle_A |-\ell\rangle_B + |-\ell\rangle_A |\ell\rangle_B \right)$ and $\rho_d$ is our measured reconstructed state. We measured a fidelity of $F = 0.98 \pm 0.01$, indicating that our reconstructed state can be considered a pure ($F = 1$), maximally entangled state.

We chose to study OAM entanglement with the objective of measuring entanglement in higher dimensions. The above mentioned density matrix was reconstructed using two bits of information per photon, that is $|\ell\rangle = \pm 1$ and its superpositions. However, in high-dimensional entanglement we encode multiple bits of information per photon. For example, for dimension $d = 3$, we could use $|\ell\rangle = 0, \pm 1$ or $|\ell\rangle = 0, \pm 5$ and for dimension $d = 5$, we could use $|\ell\rangle = -2, -1, 0, 1, 2$. An equivalent density matrix to Eq. (20) can be written for $d$-dimensions as,

$$\rho = \sum_{m,n}^{d^2-1} b_{m,n} \tau_m \otimes \tau_n , \qquad (25)$$

where $b_{m,n}$ are complex coefficients with $b_{0,0} = 1/d^2$ for normalisation. The Gell-Mann matrices[39] are represented by $\tau_q$ for $q = 1...(d^2 - 1)$ and $\tau_0$ is the $d$-dimensional identity matrix. Thus, following a similar procedure used for $d = 2$, we reconstructed the density matrix for higher dimensions from $d = 3$ to $d = 5$. A detailed description on similar measurements can be found in Agnew et al.[40] The linear entropy and fidelity of each density matrix was calculated and plotted in Fig. (16).

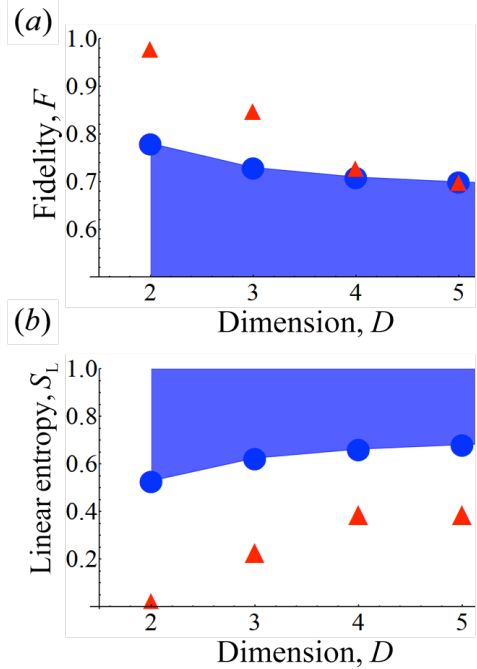

Fig. 16: (a) Fidelity and (b) linear entropy as a function of dimension. The red triangles represent the experimental measurements, while the blue dots represent the threshold states in Eq. (26).

The results indicate that the degree of entanglement decreases as the dimensionality increases, however the results do not fall below the threshold states, which lie on the threshold of the high-dimensional Bell inequality[41]. These are given by

$$\rho_B = p_d^{\min}|\psi\rangle\langle\psi| + \left(1 - p_d^{\min}\right)\frac{I}{d^2},  \qquad (26)$$

where $p_d^{\min}$ is the probability above which the Bell inequality is violated, $|\psi\rangle$ is the maximally entangled state of two $d$-dimensional systems and $I$ is the identity matrix of dimension $d^2$.

## Discussion

We have demonstrated a method to achieve a quantum entanglement setup together with procedures to measure and quantify the entangled system. By using the OAM basis we can choose to perform measurements in two-dimensional entangled states, or extend the setup to higher dimensions by altering the phase holograms used in the measurement scheme[40]. For example, we have shown a violation of a Bell-type inequality for dimension two and high fidelity states in dimension five (Hilbert space of dimension twenty five). High-dimensional entanglement offers further possibilities such as closing the detection loophole in Bell experiments[42]. Likewise, instead of using a Gaussian shaped source to pump the crystal, any number of pump shapes can be used, for example a Hermite-Gaussian beam[43]. In so doing, the down-conversion process can be better understood, possibly improving the measured spiral bandwidth.